\documentstyle[12pt]{article}

\large
\newcommand{\be}{\begin{eqnarray}}
\newcommand{\ee}{\end{eqnarray}}
\newcommand\del{\partial}

\begin{document}
\setlength{\baselineskip}{21pt}
\pagestyle{empty}  
\vfill                                                                          
\eject                                                                          
\begin{flushright}                                                              
SUNY-NTG-96/1
\end{flushright}                                                                
                                                                                
\vskip 2.0cm 
\centerline{\Large  Universality near zero virtuality}
\vskip 2.0 cm                                                                   
\centerline{\bf A.D. Jackson, M.K. \c Sener and J.J.M. Verbaarschot}
\vskip 0.2cm            
\centerline{Department of Physics}
\centerline{SUNY, Stony Brook, New York 11794}
\vskip 2cm                                                                   
                                                                                
\centerline{\bf Abstract}

In this paper we study a random matrix model with the chiral and flavor 
structure of the QCD Dirac operator and a temperature dependence given 
by the lowest Matsubara frequency. Using the supersymmetric method for random
matrix theory, we obtain an exact, analytic expression for the 
average spectral density. In the large-$n$ limit, 
the spectral density can be obtained from the solution to a cubic 
equation.  This spectral density is non-zero in the vicinity of eigenvalue 
zero only for temperatures below the critical temperature of this model.
Our main result is the demonstration that the microscopic limit of 
the spectral density is independent of temperature up to the critical 
temperature.  This is due to a number of `miraculous' cancellations.  This 
result provides strong support for 
the conjecture that the microscopic spectral density is universal. 
In our derivation, we emphasize the symmetries of the partition function and
show that this universal behavior is closely related to the existence of
an invariant saddle-point manifold.

\vfill                                                                          
\noindent                                                                       
\begin{flushleft}
January 1996
\end{flushleft}
\eject
\pagestyle{plain}

\section{Introduction}

Since its initial application to level correlations in nuclear 
spectra \cite{Porter},
random matrix theory has been applied to a variety of physical phenomena 
ranging from resonant cavities \cite{Koch} to lattice gauge theory \cite{HV}. 
In particular, the spectra of classically chaotic systems have been analyzed 
in great detail within this framework \cite{Bohigas,Berry,Altshuler}.  
Recently, 
a great deal of progress has been made in the realm of mesoscopic physics.  
For example, universal conductance fluctuations have been understood 
thoroughly within the framework of random matrix 
theory \cite{Mello,Iida,Slevin-Nagao,Beenhakker}.

In initial applications, random matrix theories were introduced in the hope of 
representing the complicated strong interactions of nuclear physics.  It 
was soon realized that the average spectral density cannot be described 
by means of random matrix theories.  (For a large class of invariant random 
matrix theories, the spectral density is given by a semicircle.  In contrast, 
the spectral density increases with energy for typical physical systems.)  
However, random matrix theories proved themselves capable of providing a 
remarkably accurate description of the correlations between eigenvalues on 
the scale of the average level spacing.  Apparently, some of the properties 
of random matrices are universal while others are not.  Such behavior is 
familiar from the theory of critical phenomena where, for example, critical 
exponents are universal, but the critical temperature is not.  There, universal 
phenomena are usually associated with the soft modes which arise due to the 
spontaneous breaking of a symmetry; non-universal properties are determined by
all modes.   A similar separation of scales takes place in random matrix 
theories.  Consider, for example, the average spectral density (with 
variations only over many level spacings) and level correlations on the 
scale of the average level spacing.  

In the supersymmetric formulation of random matrix theory, universal 
properties, {\em e.g.}, the level correlations, are associated with the 
existence of a saddle-point manifold which is intimately related to the 
symmetries of the theory.  Non-universal properties, such as the average 
spectral density, can be calculated by a saddle-point approximation. 

In this paper we study a model which was introduced in \cite{JV,Tilo}.  
This model
is a random matrix model which possesses the chiral and flavor structure of 
the QCD Dirac operator and a schematic temperature dependence corresponding 
to the lowest Matsubara frequency.  Otherwise, all matrix elements of the 
Dirac operator are completely random.  The temperature dependence is such 
that this model has a second-order phase transition with mean field critical 
exponents.  Below the critical temperature, chiral symmetry is broken
spontaneously; above the critical temperature, it is restored.  According
to the Banks-Casher formula \cite{Banks-Casher}, the order parameter is the 
spectral density at eigenvalue zero.  Physical motivation for this model 
comes from two rather different directions.  First, at zero temperature, it 
satisfies all Leutwyler-Smilga sum-rules \cite{LS}, which are identities 
for chiral QCD in a finite volume.  Second, Koci{\'c} and Kogut have recently 
suggested \cite{Kocic-Kogut} that the chiral phase transition in fermionic 
systems is driven towards a mean field description because of the fact that 
the lowest Matsubara frequency non-zero.  

The chiral structure of the Dirac operator forces all eigenvalues to appear 
in pairs $\pm \lambda$.  The spectrum is symmetric about zero.  As we shall 
soon see, it is useful to introduce the microscopic limit of the spectral 
density which probes the spectrum around zero 
on a scale set by the distance between
adjacent eigenvalues:
\be
\rho_S(u) = \lim_{N\rightarrow \infty} \frac 1{N\Xi}\rho(\frac u{N\Xi}) \ \ .
\label{microscopic}
\ee
Here, $\Xi$ is the temperature-dependent chiral condensate which, according 
to the Banks-Casher formula \cite{Banks-Casher}, is given by 
\be
 \Xi =\frac{\pi\rho(0)}{N}\ \ ,
\label{BC}
\ee
and $N$ is the total number of eigenvalues.

The zero-temperature version of this model has been studied extensively 
in the literature 
\cite{FOX-KAHN-1964,BRONK-1965,WIDOM-TRACY-1992,Forrester-Nagao,Nagao,Zee}. 
It is known as the Laguerre ensemble or the chiral Gaussian Unitary Ensemble 
(chGUE).  Two types of universal behaviour are known to exist in chiral 
random matrix theories:  Spectral correlations in the bulk of the 
spectrum are universal; the microscopic limit of the spectral density, just 
introduced, is universal.  It has been shown 
\cite{bulk,Nagao} that the chiral structure of the random matrix ensemble 
does not affect eigenvalue correlations in the bulk of the spectrum.  
Such level correlations have been observed both experimentally and numerically
in many systems \cite{Porter,Koch,Bohigas,HV}. Further, analytic arguments 
have been presented \cite{Berry,Altshuler} in favor of the universality 
of correlations in the bulk of the spectrum of classically chaotic systems. 

In \cite{SVZr} we conjectured that the microscopic limit of the 
spectral density is
universal as well.  The first argument in support of this conjecture came 
from instanton liquid calculations \cite{V2}, where we were able to generate 
ensembles large enough to permit the calculation of the spectral 
density in the microscopic limit.  A slightly 
less direct argument came from lattice QCD calculations 
of the dependence of the chiral condensate on the valence quark mass 
\cite{Christ,Vcr}.  
Another hint came from the work of the MIT group \cite{MIT}, 
who studied the chGUE using the supersymmetric method.  They found that the 
microscopic limit of the 
spectral density is determined by a saddle-point manifold 
associated with the spontaneous breaking of a symmetry.  The first convincing 
analytical arguments in favor of this conjecture came from recent work by 
Br{\'e}zin, Hikami and Zee \cite{Zee}.  They considered families of random 
matrix models all possessing the chiral structure of the Dirac operator.  
They discovered the same microscopic limit in all the models 
they investigated.  

In this paper, we offer further evidence in support of the universality of 
the microscopic limit of the 
spectral density.  Specifically, we investigate the effect 
of temperature in the chiral random matrix model introduced in \cite{JV,Tilo}. 
This model differs structurally from the models in \cite{Zee}.  
In the random matrix models considered in \cite{Zee}, the unitary symmetry
of the probability distribution leaves the spectrum of each element of the
ensemble invariant.
This invariance 
is not realized for temperatures $T \ne 0$, 
and analytic proofs are consqeuently somewhat 
more difficult.  Using the supersymmetric
method of random matrix theory, we obtain an exact expression for
the spectral density which is valid for any dimension, $n$, of the matrices.  
This enables us to take the microscopic limit.  This limit also requires the 
large-$n$ limit of the spectral density (see (\ref{BC})), which can be 
evaluated conveniently by means of a saddle-point approximation.  In \cite{JV}, 
this spectral density was evaluated numerically.  It was found to have 
the well-known semi-circular shape at zero temperature.  At high temperature, 
the shape is given by two disjoint semi-circles with centers located at 
$\pm \pi T$.  In that paper, we also announced the analytic result for 
the shape of the average spectral density. The result 
merely requires the solution of a cubic equation.  This result 
has also been obtained by Stephanov \cite{Stephanov}, who applied
an extension of this model to the problem of the relation between the
$Z_N$ phase of the theory and the restoration of chiral symmetry.
 
The organization of this paper is a follows. In section 2 we give a definition
of the random matrix model and the supersymmetric partition function. In 
section 3 the partition function is reduced to a finite-dimensional integral.
Symmetries and convergence questions of the partition function are analyzed
in section 4. The exact two-dimensional integral for the resolvent is
obtained in section 5. In section 6 we derive the large-$n$ limit of the
spectral density and discuss its properties. The microscopic limit of the
partition function is evaluated in section 8, and concluding remarks are
made in section 9.  Our notation and conventions are explained in
Appendix A, and a perturbative calculation of the large-$n$ limit of the
spectral density is presented in Appendix B.

\vskip 1.5 cm
\section{Definition of the random matrix model}
\vskip 0.5cm

In this paper we study the spectrum of the ensemble of matrices
\be
H = \left (
\begin{array}{cc} 0 & W+\pi T \\ W^\dagger +\pi T & 0
\end{array} \right ) \ \ .
\label{matrix}
\ee
Here, $T$ is the temperature dependence as given by the lowest Matsubara 
frequency, and $W$ is a complex $n\times n$ matrix distributed according to
\be
 \exp[-{n\Sigma^2} \,{\rm Tr}\, W W^\dagger] \ \ .
\label{probability}
\ee
The average spectral density can be expressed as
\be
\rho(\lambda) = -\lim_{\epsilon \rightarrow 0}
\frac {2n}{\pi} \ {\rm Im} \ G(\lambda + i\epsilon) \ \ .
\label{spectrum}
\ee 
where the average resolvent $G(z)$, 
\be
G(z) = \frac 1{2n}{\rm Tr}\overline{\frac 1{z+i0-H}}
\ = \ - \left. \frac 1{2n}\frac {\del \log Z(J)}{\del J}\right |_{J=0} \ \ ,
\label{resolvent}
\ee
can be obtained from the partition function 
\be
Z(J) = \int {\cal D} W \ \frac{{\det}(z-H)}{\det(z+J -H)}
\ \exp[- n \Sigma^2 \,{\rm Tr}\, W W^\dagger]. \ \nonumber\\
\label{zrandom}
\ee
The integration measure, ${\cal D} W$, is the Haar measure normalized so 
that $Z(0) = 1$.  For a Hermitean matrix, $H$, the resolvent is analytic 
in $z$ in the upper complex half-plane. This allows us to calculate the 
resolvent for purely imaginary $z$ and to perform the analytic continuation 
to real $z$ at the end of the calculation.  As will be seen below, this 
improves the convergence properties of the integrals in the partition function. 

Some properties of this model all already known.  At $T=0$, this model 
reduces to the well-known Laguerre ensemble.  The joint probability 
distribution of the eigenvalues is known explicitly, as are all correlation 
functions.  In particular, the average spectral density is a semicircle: 
\be
{\rho}(\lambda) = \frac{n\Sigma^2}\pi \sqrt{\frac 4{\Sigma^2}-\lambda^2} \ \ .
\label{semicircle}
\ee
We wish to stress that the largest eigenvalue is larger than a typical 
matrix element by a factor on the order of $\sqrt n$.  

The temperature dependence of this model was analyzed in \cite{JV}. It was 
shown that, in the thermodynamic limit, this model shows a chiral phase 
transition at a critical temperature of
\be
T_c = \frac 1{\pi \Sigma} \ \ .
\ee
The order parameter is the chiral condensate $\Xi$ with $\Xi \ne$ below 
$T_c$ and $\Xi = 0$ above $T_c$.  This chiral symmetry is broken 
spontaneously.  For each finite value of $n$, $\Xi = 0$.  A non-zero 
value of $\Xi$ is obtained only in the thermodynamic limit.
Below $T_c$ it was found that in this limit
\be
\Xi = \Sigma \ (1 - \pi^2 T^2 \Sigma^2 )^{1/2} \ \ .
\label{chicon}
\ee
The 
Banks-Casher formula (\ref{BC}) allows us to convert this value of $\Xi$ 
into the spectral density $\rho(0)$.
In \cite{JV}, the complete spectral density of this model was determined 
numerically.  At $T=0$ the result (\ref{semicircle}) was reproduced;
at $T=T_c$ we found that $\rho(\lambda) \sim \lambda^{1/3}$.  For $T \gg T_c$, 
the spectral density reduced to two semi-circles centered at $\pm \pi T$ 
with a radius independent of $T$.  We will present an analytic derivation 
of these results.
\vskip 1.5cm
\section{Ensemble average of the partition function}
\vskip 0.5cm
In order to perform the Gaussian integrals, we write the determinant as 
an integral over the fermionic variables $\chi$ and $\chi^*$:
\be
\det(z - H) & = & (2 \pi)^{2n}\int \prod_{i=1}^n d[{\chi_1}_i^*] d[{\chi_1}_i]
\prod_{i=1}^n d[{\chi_2}_i^*] d[{\chi_2}_i] 
\nonumber \\
& {} & \qquad \times \exp i
\left ( \begin{array}{c} \chi_1^* \\ \chi_2^* \end{array} \right )
\left ( \begin{array}{cc} z & - W- \pi T\\ -W^\dagger -\pi T &z 
\end{array} \right ) 
\left ( \begin{array}{c} \chi_1 \\ \chi_2 \end{array} \right ) \ \ .
\label{zfermion}
\ee
Similarly, the inverse determinant can be written as an integral over the 
bosonic variables $\phi$ and $\phi^*$:
\be
{\det}^{-1}(z - H) & = & \frac{1}{(2 \pi)^{2n}}\ \int 
\prod_{i=1}^n d[{\phi_1}_i] d[{\phi_1}_i^*] 
\prod_{i=1}^n d[{\phi_2}_i] d[{\phi_2}_i^*] 
\nonumber \\
& {} & \qquad \times \exp i
\left ( \begin{array}{c} \phi_1^* \\ \phi_2^* \end{array} \right )
\left ( \begin{array}{cc} z +J& - W- \pi T\\ -W^\dagger -\pi T &z+J 
\end{array} \right ) 
\left ( \begin{array}{c} \phi_1 \\ \phi_2 \end{array} \right ) \ \ .
\label{zboson}
\ee
The conventions for the Gaussian integrals are defined in Appendix A.
The factor $i$ in the exponent in (\ref{zboson}) is chosen so that the
integral is convergent for $z$ in the upper complex half-plane. This choice
is consistent with the $i \epsilon$ prescription in (\ref{resolvent}).
The integral in (\ref{zfermion}) converges independent of the overall
phase of the exponent.  The present choice of phase ensures that the product 
of the fermionic and bosonic integrals is one.

The Gaussian integral over $W$ can be performed by completing the squares. The 
result is a term of fourth order in the integration variables,
\be
\exp -\frac 1{n\Sigma^2} (\chi_{2j}^* \chi_{1i} + \phi_{2j}^* \phi_{1i})
(\chi_{1i}^* \chi_{2j} + \phi_{1i}^* \phi_{2j}) \ \ .
\ee
We apply the Hubbard-Stratonovitch transformation to each of the terms of 
fourth order in the bosonic and fermionic variables.  Two of the four 
factors can be decoupled with the help of real integration variables: 
\be
\exp- \frac 1{n\Sigma^2} \phi_1^*\cdot \phi_1 
\, \phi_2^*\cdot \phi_2
= \int \frac{d\sigma_1 d\sigma_2}{I_b} 
\exp[ -n\Sigma^2(\sigma_1^2 + \sigma_2^2)
-(\sigma_1+i\sigma_2) \phi_1^*\cdot \phi_1
+(\sigma_1-i\sigma_2) \phi_2^*\cdot \phi_2 ] \ , \nonumber \\
\exp+ \frac 1{n\Sigma^2} \chi_1^*\cdot \chi_1 \, \chi_2^*\cdot \chi_2
= \int \frac{d\rho_1 d\rho_2}{I_b} \exp[ -n\Sigma^2(\rho_1^2 + \rho_2^2)
-(\rho_1-i\rho_2)\chi_1^*\cdot \chi_1
-(\rho_1+i\rho_2)\chi_2^*\cdot \chi_2 ]\ . \nonumber \\
\label{rhosig}
\ee
The terms which involve mixed bilinears can be decoupled with the help 
of Grassmann integrations.  We do not encounter convergence problems 
in the process.  
\be
\exp- \frac 1{n\Sigma^2} \chi_1^*\cdot \phi_1 \, \chi_2\cdot \phi_2^*
&=& I_f \int \frac{d\alpha^* d\beta}{(i/2)} \exp[ -n\Sigma^2\alpha^* \beta
+\alpha^*\chi_1^*\cdot \phi_1
-\beta\phi_2^*\cdot \chi_2 ] \ , \nonumber \\
\exp \frac 1{n\Sigma^2} \chi_1\cdot \phi_1^* \, \chi_2^*\cdot \phi_2
&=& I_f \int \frac{d\beta^* d\alpha}{(i/2)} \exp[- n \Sigma^2 \beta^* \alpha
-\alpha \chi_1 \cdot \phi_1^*
+\beta^* \phi_2 \cdot \chi_2^* ] \ \ .
\label{after-averaging}
\ee
The constants $I_b$ and $I_f$ are defined such that $I_f/ I_b = 1$.
(See Appendix A.) 

Thus, we obtain the partition function as 
\be
Z(J) &=& \int \prod_{i=1}^{n} d[{\phi}_i ] \ d[{\chi}_i ] \ d[\sigma]
\exp \left[ -n\Sigma^2(\sigma_1^2+\sigma_2^2
+\rho_1^2 + \rho_2 ^2 + \alpha^* \beta + \beta^* \alpha) \right]\nonumber\\
&\times&
\exp i\left (
\begin{array}{c} \phi_1^* \\ \phi_2^* \\ \chi_1^* \\ \chi_2^* 
\end{array} \right )
\left ( \begin{array}{cccc}
           z+J+i\sigma_1 -\sigma_2 & -\pi T & i\alpha & 0 \\
             -\pi T  & z+J -i\sigma_1 -\sigma_2 & 0 & i\beta \\
            i\alpha^* & 0& z+i\rho_1+\rho_2 &-\pi T\\
             0  & i\beta^* & -\pi T & z  +i\rho_1 -\rho_2 
\end{array} \right )
\left ( \begin{array}{c} \phi_1 \\ \phi_2 \\ \chi_1 \\ \chi_2 
\end{array} \right ),\nonumber \\
\label{voorlaatst}
\ee
where
\be
d[\sigma] =
d\sigma_1 \ d\sigma_2 \ d\rho_1 \ d\rho_2 \ 
\frac{ d\alpha \ d\alpha^* \ d\beta \ d\beta^*}{(i/2)^2}\ . 
\ee
If $i \sigma_1$, $\sigma_2$, $z$, and $J$ are 
all real, the matrix $A$ appearing in the exponent of (\ref{voorlaatst})
is a graded Hermitean matrix. Then 
the Gaussian integrals can be performed according to (\ref{supergauss}).
This results in
\be
Z(J) = \int d[\sigma]
\exp \left [ -n\Sigma^2(\sigma_1^2+\sigma_2^2
+\rho_1^2 + \rho_2 ^2 + \alpha^* \beta + \beta^* \alpha) \right ]\,\,
{\rm detg}^{-n} A \ ,
\label{finalz}
\ee
where ${\rm detg } A$ is the graded determinant of $A$. 
For a matrix with 
Grassmann blocks $\rho$ and $\sigma$ and commuting blocks $a$ and $b$, 
it can be shown that
\be
{\rm detg} \left ( \begin{array}{cc} a& \sigma \\ \rho & b \end{array}\right)
= {\det}^{-1} b \ \det(a - \sigma b^{-1} \rho) \ .
\label{detg}
\ee
In our case, all blocks $a$ and $b$ are $2 \times 2$ matrices, which 
permits us to evaluate all expressions directly.

We note, however, that the result (\ref{finalz}) was obtained by
interchanging the $\phi_i$ and $\sigma_i$ integrations in (\ref{voorlaatst}).  
This is allowed only if the $\phi$ integral is uniformly convergent in 
$\sigma$.  Unfortunately, this is not the case when the $\sigma_1$ and 
$\sigma_2$ integration paths are along the real axis.  This problem can 
be circumvented by a suitable deformation of the integration paths.  
Previous studies of random matrix theories within the framework of the
sigma model formulation of the Anderson model 
\cite{sigma} have stressed the importance of deforming the 
integration contours in a manner which is consistent with the symmetries 
of the problem.  The same is true for the present problem.  In order to 
interchange the $\phi$ and $\sigma$ integrals in (\ref{voorlaatst}), we 
must deform the integration contour so that the $\phi$ integration is 
uniformly convergent in $\sigma$.  In order to motivate our choice of 
contour, we must first consider the symmetries of the 
partition function. 

\vskip 1.5cm
\section{Symmetries}
\vskip 0.5cm

We wish to study the partition function in the microscopic limit, {\em i.e.},  
the limit $n \rightarrow \infty$ with $zn$ held constant.  For $z = 0$ (and
$J= 0$), the partition function has an additional symmetry.
For all temperatures, the bosonic part of the partition function 
is invariant under the {\em non-compact\/} symmetry operation
\be
\phi_1 \rightarrow e^{+t} \phi_1 \ , \qquad \phi_1^* \rightarrow e^{+t} 
\phi_1^* \ ,\nonumber\\ 
\phi_2 \rightarrow e^{-t} \phi_2 \ , \qquad \phi_2^* \rightarrow e^{-t} 
\phi_2^* \ . 
\label{noncompact}
\ee
This induces a hyperbolic rotation of the variables $\sigma_1$ and 
$\sigma_2$ of the preceding section, 
\be
\left ( \begin{array}{c} \sigma_1 \\ \sigma_2 \end{array} \right ) \rightarrow
\left (\begin{array}{cc} \cosh t & i\sinh t\\ 
                        -i \sinh t & \cosh t \end{array} \right )
\left ( \begin{array}{c} \sigma_1 \\ \sigma_2 \end{array} \right ) \ ,
\ee
which clearly reveals the $O(1,1)$ nature of the transformation.

The fermionic part of the partition function is invariant under
\be
\chi_1 \rightarrow e^{+iu} \chi_1 \ , \qquad \chi_1^* \rightarrow e^{+iu} 
\chi_1^* \ ,\nonumber\\ 
\chi_2 \rightarrow e^{-iu} \chi_2 \ , \qquad \chi_2^* \rightarrow 
e^{-iu}\chi_2^* \ , 
\ee
where, {\em a priori}, $u$ can be either real or complex.  In terms of the 
$\rho$ variables of (\ref{rhosig}), this induces the transformation 
\be
\left ( \begin{array}{c} \rho_1 \\ \rho_2 \end{array} \right ) \rightarrow
\left (\begin{array}{cc} \cos u & \sin u\\ 
                        - \sin u & \cos u \end{array} \right )
\left ( \begin{array}{c} \rho_1 \\ \rho_2 \end{array} \right ).
\label{compact}
\ee
Because the integration over the Grassmann variables is finite, the volume
of the symmetry group must be finite as well.  Therefore, $u$ must be 
real with $u \in [0, 2\pi]$.  In other words, the symmetry group is $O(2)$.

The terms of a mixed fermionic-bosonic nature are also affected by 
the transformations (\ref{noncompact}) and (\ref{compact}).  This induces 
a transformation of Grassmann variables introduced through the 
Hubbard-Stratonovitch transformation.  

As is known from studies of random matrix theories, the 
parameterization of the variables $(\sigma_1, \sigma_2)$ and 
$(\rho_1,\rho_2)$ is dictated by the above symmetries.  It is natural 
and convenient to choose 
integration variables which lie along and perpendicular to the invariant 
manifold.  In the microscopic limit, integrations along this manifold must 
be performed exactly, whereas the perpendicular integrations can be performed 
by saddle-point methods in the limit $n \rightarrow \infty$.
For the $\sigma$ variables, we thus choose the parametrization
\be
\sigma_1 &=& -i(\sigma-i\epsilon) \sinh s/\Sigma \ , \nonumber\\
\sigma_2 &=&  (z - i \epsilon) + J +(\sigma-i\epsilon) \cosh s/\Sigma \ ,
\label{parbb}
\ee
where $\sigma \in [-\infty; +\infty]$ and $s \in [-\infty;+\infty]$.  
After the $\phi_i$ integration, the $i\epsilon$ appears only in the 
combination $\sigma -i\epsilon$.  Below, we will not write the $i\epsilon$ 
term explicitly, but it is always understood that it is included in the 
variable $\sigma$.  This parametrization renders the $\phi_1$ and $\phi_2$ 
integrations uniformly convergent in $\sigma_1$ without jeopardizing the 
convergence of the $\sigma$ and $s$ integrations.  This allows 
us to interchange the $\phi_i$ and 
$\sigma_i$ integrations leading to the final result (\ref{finalz}) of the 
last section.  The term $\sigma_1^2 +\sigma_2^2$ appearing in the first 
exponent in (\ref{finalz}) becomes $\sigma^2 + (z+J)^2 +2(z+J)\sigma\cosh s$ 
in the parametrization (\ref{parbb}).  It is clear that the integral over 
$s$ can be convergent only when $z+J$ is purely imaginary.  (Recall that 
$\sigma$ contains the term $-i \epsilon$.)  The transformation 
(\ref{noncompact}) for $z+J=0$ 
reduces to a translation of $s$ and leaves $\sigma$ invariant.

The $\rho$ variables are also parametrized along and perpendicular to the 
saddle-point manifold according to
\be
\rho_1 &=&iz+ \rho \cos \varphi/\Sigma \ , \nonumber \\
\rho_2 &=& \rho \sin \varphi/\Sigma \ .
\label{parff}
\ee
The rotation (\ref{compact}) leads to a translation of the angle $\varphi$ and 
leaves $\rho$ invariant.

Our strategy in dealing with (\ref{finalz}) is to perform the Grassmann 
integrations first, {\em i.e.}, to collect the coefficient of $\alpha 
\alpha^* \beta \beta^*$.  This leaves
us with a four-dimensional integral which is the exact analytical result
for the partition function for any finite $n$.  In the thermodynamic limit and
for $z \sim {\cal O}(1)$, the remaining integrations can be performed with 
a saddle-point approximation.  This result is obtained in section 6.
In the microscopic limit, $z$ will be ${\cal O}(1/n)$, and the integration 
over the invariant manifold must be performed exactly.  The radial 
integrals can be approximated to leading order in $1/n$. (See section 5.)

The $T=0$ problem has been investigated previously using the supersymmetric 
method \cite{MIT,Zee}.  In \cite{MIT}, the saddle-point manifold was 
constructed in a manner similar to that used for the problem of invariant 
random matrix ensembles.  (In this regard, see \cite{Efetov,SVZ}.)  In 
\cite{Zee}, the convergence difficulties were circumvented in an elegant 
fashion by the use of spherical coordinates for the variables $\phi_1$ and 
$\phi_2$.  Unfortunately, a direct generalization of this approach is not 
possible for the present case of non-zero temperatures.  When $T \ne 0$, 
the angles between the complex vectors $\phi_1$ and $\phi_2$ also enter 
in the integration variables.

\vskip 1.5cm
\section{Exact result for the spectral density at finite $n$}
\vskip 0.5cm

Because the fermionic blocks of the matrix $A$ are nilpotent, the right 
side of (\ref{detg}) can be expanded in a finite number of terms.
The $n$-th power of the inverse of the graded determinant can be written as 
\be
{\rm detg}^{-n} 
\left ( \begin{array}{cc} a& \sigma \\ \rho & b \end{array}\right)
= \left( \frac{\det b}{ \det a} \right )^{n}\left ( 1 + 
n {\rm Tr} \, a^{-1} \sigma 
b^{-1} \rho + \frac n2 {\rm Tr}\, (a^{-1} \sigma b^{-1}\rho )^2 +\frac{n^2}2
{\rm Tr}^2 \, a^{-1} \sigma b^{-1}\rho \right ) \ . \nonumber\\
\ee
Terms in the partition function which are of fourth order in the Grassmann 
variables can be obtained by supplementing the above terms with factors 
$\alpha^* \beta$ and $\beta^* \alpha$ from the exponent in (\ref{finalz}). 
The result is
\be
Z(J) &=& \frac {n^2\Sigma^4}{\pi^2}
\int d\sigma_1 d\sigma_2 d\rho_1 d\rho_2 \left 
[(1-\frac{\pi^2 T^2}{D\Delta\Sigma^2})^2 - \frac {(D+ \pi^2 T^2)(\Delta + \pi^2 
T^2)+\pi^2T^2(\Delta-D)/n }{D^2\Delta^2\Sigma^4}
\right ]\nonumber\\
&\times& 
\left ( \frac \Delta D \right )^n \exp [-n\Sigma^2
(\sigma_1^2+\sigma_2^2+ \rho_1^2 + \rho_2^2)] \ ,
\label{zint}
\ee
where $D$ is the determinant of the boson-boson block,
\be
D = (z+J+i\sigma_1 -\sigma_2)(z+J -i\sigma_1 -\sigma_2) -\pi^2 T^2 \ ,
\ee
and $\Delta$ is the determinant of the fermion-fermion block,
\be
\Delta =(z+i\rho_1+\rho_2)(z  +i\rho_1 -\rho_2) -\pi^2 T^2 \ .
\ee

In (\ref{zint}) the variables $\sigma_i$ and $\rho_i$ are parametrized
according to (\ref{parbb}) and $(\ref{parff})$.
temperature by
\be
t = \pi T \Sigma \ 
\ee
the resulting form of $Z(J)$ simplifies to 
\be
Z(J) &=& \frac {-in^2}{\pi^2}
\int_{-\infty}^{\infty} \sigma d\sigma \int_{-\infty}^{\infty} ds
\int_0^{\infty}\rho d\rho \int_0^{2\pi} d\varphi\nonumber \\
&\times& \left
[(1-\frac{t^2}{(t^2-\sigma^2)(\rho^2+t^2)})^2 + 
\frac {\sigma^2\rho^2+t^2(\sigma^2+\rho^2)/n }
{(\sigma^2-t^2)^2(\rho^2+t^2)^2}
 \right ] \nonumber \\
&\times&
\left ( \frac {\rho^2 +t^2}{t^2-\sigma^2}
 \right )^n e^{-n(\sigma^2
+ \rho^2 +2(z+J)\Sigma\sigma \cosh s + 2iz\Sigma\rho \cos \varphi 
+\Sigma^2((z+J)^2 -z^2))} \ .
\label{lastreal}
\ee
The integrations over $s$ and $\varphi$ can be expressed in terms of Bessel
functions.  Differentiation of the partition function with respect to $J$ 
at $J=0$ gives us the resolvent which we desire.  Thus, the final result of 
this section is 
\be
G(z) &=& \frac {2in}{\pi}
\int  \sigma d\sigma \rho d\rho \nonumber \\
&\times&\left
[(1-\frac{t^2}{(t^2-\sigma^2)(\rho^2+t^2)})^2 + 
\frac {\sigma^2\rho^2+t^2(\sigma^2+\rho^2)/n }
{(\sigma^2-t^2)^2(\rho^2+t^2)^2}
 \right ] \nonumber \\
&\times&
\left ( \frac {\rho^2 +t^2}{t^2-\sigma^2}
 \right )^n
 (2z\Sigma^2 K_0(2n\Sigma z \sigma) -2n\Sigma\sigma K_1(2n\Sigma z \sigma))
 J_0(2n\Sigma z\rho)
\exp[-n(\sigma^2+ \rho^2)]. \nonumber\\
\label{finalg}
\ee
Again, we remind the reader that $\sigma$ contains a term $-i\epsilon$. 
The integral over $\sigma$ can thus be performed by successive partial 
integrations.  Details regarding the calculation of this kind of integral 
can be found in \cite{Guhr1}.  This result has been obtained for $z$ purely
imaginary.  Since the modified Bessel functions have a cut for $z \sigma < 0$,
we can analytically continue this expression anywhere in the upper half-plane.

\vskip 1.5cm
\section{The large-$n$ limit of the average spectral density}
\vskip 0.5cm

For $n \rightarrow \infty $ and $z\sim {\cal O}(1)$, all integrals in 
(\ref{lastreal}) can be performed by a saddle-point approximation.  
Because we started 
with a supersymmetric partition function, the Gaussian fluctuations
about the saddle point give an overall constant of unity, {\em i.e.}, 
$Z(0) = 1$ in (\ref{lastreal}).  Using the relation (\ref{resolvent}) 
to determine the resolvent from the partition function (\ref{lastreal}), 
we find that 
\be
G(z) = \Sigma (\Sigma z +\bar \sigma \cosh \bar s) \ ,
\ee 
where $\bar \sigma$ and $\bar s$ are the saddle-point values of these
variables.  The saddle-point equation for $s$ is trivial with solution 
$\bar s = 0$.  The equation for $\bar \sigma$ is more complicated
\be
\frac {\bar \sigma}{\bar \sigma^2-t^2} - (\bar\sigma + \Sigma z) = 0 \ .
\label{spe}
\ee
This equation can be rewritten as an equation for the ensemble averaged 
resolvent
\be
G^3/\Sigma^4 - 2z G^2 /\Sigma^2 + G(z^2 -\pi^2 T^2 + 1/\Sigma^2) -z = 0 \ .
\label{speg}
\ee
At $T=0$, this equation reduces to
\be
(G-z)(G^2/\Sigma^2-zG +1) = 0
\ee
with a non-trivial solution 
\be
G(z) = \Sigma^2 \ \frac{z \pm i( 4/\Sigma^2-z^2)^{1/2}}{2} \ .
\ee
As indicated in (\ref{spectrum}) above, the associated spectral density 
is simply the imaginary part of the branch of $G(z)$ with the negative 
sign, 
\be
\rho (\lambda) = \frac {n\Sigma^2}\pi ( 4/\Sigma^2-\lambda^2)^{1/2} \ ,
\ee
which is the familiar semicircle normalized to the total number of eigenvalues.

For $z = 0$, the saddle-point equation (\ref{speg}) simplifies to 
\be 
G^3 + \Sigma^4 G(1/\Sigma^2 - \pi^2T^2) = 0
\ee
with the solution
\be 
G(0) = -i \Sigma\sqrt{1-\pi^2 T^2 \Sigma^2} \ .
\ee
This corresponds to the spectral density 
\be
\rho(0) = \frac {2n\Sigma}{\pi} \sqrt{1- \pi^2 T^2 \Sigma^2} \ \ .
\label{rho(0)}
\ee
Using the Banks-Casher formula (\ref{BC}), we immediately obtain the chiral
condensate (\ref{chicon}) in agreement with \cite{JV}.

In order to determine the high-temperature limit of the spectral density,
it is most convenient to return to the saddle-point equation (\ref{spe}). 
It is clear that, for $z\approx \pi T$, this equation can only be satisfied 
for $\bar \sigma \approx -t$. In the partition function (\ref{lastreal})
we can approximate the logarithmic term 
\be
\log(t^2-\sigma^2) \approx \log(2t) + \log (t+\sigma) \ \ .
\ee
This leads us to the high-temperature limit of the saddle-point equation
\be
-\frac {1}{\bar \sigma+ t} - (\bar\sigma + \Sigma z) = 0 \ .
\label{spet}
\ee
The solution for the resolvent is 
\be
G(z) = \frac \Sigma 2 (\Sigma z -t -i \sqrt{2-(\Sigma z-t)^2 }) \ \ .
\label{Ghigh}
\ee
This results in a semicircular spectral density of radius $\sqrt 2$ 
located at $z = \pi T$.  An identical argument leads to another 
semicircular contribution to the spectral density of radius $\sqrt{2}$ 
centered at $z = - \pi T$.  Of course, we can arrive at the same conclusion 
working directly from (\ref{speg}).  For $z \approx \pi T$, the 
resolvent $G(z) \sim {\cal O}(1)$, 
and the first term in (\ref{speg}) will be sub-leading in the high-temperature
limit.  This leads immediately to (\ref{Ghigh}).  

Finally, we consider the case at the critical temperature, $T=\Sigma/\pi$, 
with $z$ in the 
neighborhood of $0$. Then, the saddle-point equation for $G$ reduces to
\be
G^3 = z
\ee
with solutions $(z\Sigma^4)^{1/3}$, $(z\Sigma^4)^{1/3}\exp(\pi i/ 3)$
and $(z\Sigma^4)^{1/3}\exp(2\pi i/ 3)$.  Only the last of these gives 
rise to a positive definite spectral density with
\be
\rho(\lambda) = \frac {n\Sigma\sqrt 3}{\pi} (\lambda\Sigma)^{1/3}
\ee
in agreement with the mean field critical exponent of $\delta = 3$
for this model.   

The equation for the resolvent (\ref{speg}) also enables us to obtain a 
simple recursion relation for the moments of the spectral density.  
Expanding $G(z)$ in terms of these moments, 
\be
G(z) = \sum_n \frac {M_{2n}} {z^{2n+1}} \ ,
\ee
we obtain
\be
M_{2n+2} = (T^2-\frac 1{\Sigma^2})M_{2n} + \frac 2{\Sigma^2}
\sum_{k+l=n}M_{2k}M_{2l}-\frac 1{\Sigma^4}\sum_{k+l+m=n-1}
M_{2k}M_{2l}M_{2m} \ \ .
\ee
The evident initial condition, $M_0 = 1$, immediately leads us to 
{\em et cetera}.  Without too much effort, it is possible to use standard 
combinatoric methods to find the general result for the $(2n)$-th moment,
\be
M_{2n} = \sum_{k=0}^n y^{2k} T^{2(n-k)} 
\frac 1{k+1}
\left ( \begin{array}{c} n \\ k \end{array} \right ) 
\left ( \begin{array}{c} 2n \\ k \end{array} \right ) \ \ .
\ee

\vskip 1.5cm
\section{The microscopic limit of the partition function}
\vskip 0.5cm

The `microscopic limit' denotes the investigation of the spectral density 
in the vicinity of $z= 0$ on a scale set by the average level spacing.  More 
precisely, we take the limit $n\rightarrow \infty$ while keeping $nz$ fixed, 
as indicated in (\ref{microscopic}). We start from the expression
(\ref{finalz}) for the partition function. In the thermodynamic limit, 
the $\sigma$ and $\rho$ integrations can be performed by a saddle-point 
method.  The saddle-point equations read
\be
\frac \rho{\rho^2 + t^2} -\rho = 0 \ , \nonumber \\
\frac \sigma{ t^2-\sigma^2 } -\sigma = 0\ , 
\ee
with solutions
\be
{\bar \rho}^2 = 1- t^2 \ , \nonumber\\
{\bar \sigma}^2 = t^2 - 1 \ .
\label{trivial1}
\ee
For temperatures less than the critical temperature, ${\bar \rho}$ is real 
and ${\bar \sigma}$ is purely imaginary.  The integration range of $\rho$ is 
the positive real axis. Therefore, the sign of ${\bar \rho}$ is positive.
The $\sigma$ integration ranges from $-\infty$ to $+\infty$.  In order the 
reach the $\sigma$ saddle point, we must deform the integration contour.
Because of the modified Bessel functions which appear in our expression 
(\ref{finalg}) for the resolvent, there is a cut in the complex 
$\sigma$-plane for $\sigma z$ on the negative real axis.  The cut of the 
modified Bessel function is then $i\epsilon$
above the positive real axis for negative $z$ and $i\epsilon$ above the
negative real axis for positive $z$.  Therefore, independent of the sign 
of $z$, only the saddle point with a negative imaginary part can be reached 
by a deformation of the contour.  Thus, 
\be
\bar \sigma = -i \left (1-t^2\right)^{1/2},
\ee
for $t < 1$.  

At the saddle point, the pre-exponential factor vanishes:
\be
\left( 1-\frac{t^2}{(t^2-\bar\sigma^2)(\bar\rho^2+t^2)} \right)^2 + 
\frac {\bar\sigma^2\bar\rho^2+ t^2(\bar\sigma^2+\bar\rho^2)/n}
{(t^2-\bar\sigma^2)^2(\bar\rho^2+t^2)^2}=0 \ .
\label{miracle1}
\ee
Given (\ref{trivial1}), it is trivial that this equation is satisfied 
when $t=0$.  However, the vanishing of this pre-exponential factor 
for arbitrary 
$t$ is remarkable and unexpected.  This fact is responsible for the 
`universal' behaviour of the microscopic limit of the spectral density.
As a consequence, the ${\cal O}(1/n)$ term in this factor does not contribute
to the resolvent to leading order in $1/n$.  The $zK_0$ term in the 
pre-exponent is also of subleading order ($z\sim {\cal O}(1/n))$.  This 
leads to the following result for the resolvent in the microscopic limit:
\be
G(z) &=& -\frac {4 n^2i\Sigma}{\pi}
\int  \sigma d\sigma  \rho d\rho \nonumber \\
&\times&\left
[ \left( 1-\frac{t^2}{(t^2-\sigma^2)(\rho^2+t^2)} \right)^2 + 
\frac {\sigma^2\rho^2}{(\sigma^2-t^2)^2(\rho^2+t^2)^2}
 \right ] \nonumber \\
&\times&
\sigma K_1(2n\Sigma z \sigma)  \, J_0(2n\Sigma z\rho)
\exp[-n( \sigma^2+ \rho^2+\log(t^2-\sigma^2) - \log(\rho^2 +t^2))] \ . 
\nonumber\\
\label{leadingz}
\ee
\newcommand{\dr}{\delta \rho}
\newcommand{\ds}{\delta \sigma}
In order to proceed, we make the substitution
\be
\sigma & = & {\bar \sigma} + \ds \ , \nonumber \\
\rho   & = & {\bar \rho}   + \dr
\ee
in (\ref{leadingz}) and keep only those terms which contribute to 
leading order in $1/n$, {\em i.e.}, terms through second order in 
$\dr$ and $\ds$.  The exponent in (\ref{leadingz}) then becomes 
\be
\exp(2n\bar\sigma^2 \ds^2 -2n\bar\rho^2 \dr^2) \ .
\label{exponent}
\ee
The product of the terms in square brackets in (\ref{leadingz}) and
$\sigma^2 \rho$ can be expanded as
\be
\bar \sigma^2 \bar \rho[
2\bar\rho^3\dr +2\bar\sigma^3 \ds +\bar\rho^2(-1+8t^2) \dr^2
+\bar\sigma^2(+1+8t^2)\ds^2  -\bar\sigma\bar\rho (2+8t^2)\dr\ds ] \ .
\label{expanded-preexponent}
\ee
It is clear already at this point that all temperature dependence enters 
through the scale factors ${\bar \sigma}$ and ${\bar \rho}$.  The terms
of ${\cal O}(\dr\ds)$ vanish upon integration with the exponential factor 
(\ref{exponent}).  Since $\langle \dr^2 \rangle = \langle \ds^2
\rangle$ and $\bar \sigma^2 = -\bar\rho^2$, the other terms involving 
$8t^2$ cancel as well.  To complete the calculation, we need only expand 
the Bessel functions to first order
\be
K_1(2n\Sigma z \sigma)\, J_0(2n\Sigma z\rho) = K_1 J_0
+2n\Sigma z \ds \,K_1'\,J_0
+2n\Sigma z \dr \, K_1 \,J_0'.
\ee
where the Bessel functions $K_1$ and $J_0$ and their derivatives appearing 
on the right of this equation are to be evaluated at their saddle points 
which are $2n\Sigma z\bar\sigma$ and $2n\Sigma z\bar\rho$, respectively.
In order to arrive at the final result, we form the product of this 
expression and (\ref{expanded-preexponent}), collect the coefficients of 
$\dr^2$ and $\ds^2$, and perform the Gaussian integrations over 
$\dr$ and $\ds$ according to
\be
\langle \dr^2 \rangle &=& \frac 12\frac{\sqrt{\pi}}{(2n\bar\rho^2)^{3/2}}
\frac{\sqrt{\pi}}{(-2n\bar\sigma^2)^{1/2}} \ , \nonumber \\
\langle \ds^2 \rangle &=& \frac 12\frac{\sqrt{\pi}}{(2n\bar\rho^2)^{1/2}}
\frac{\sqrt{\pi}}{(-2n\bar\sigma^2)^{3/2}} \ .
\ee
The result is
\be
G(z) = -\frac {i\Sigma}{2} \frac {\bar\sigma^2}{\bar\rho^3 }
 \left[
K_1 J_0 ({\bar \sigma}^2- {\bar \rho}^2) + 4nz{\bar \rho}^3 \Sigma K_1 J_0' 
+ 4nz {\bar \sigma}^3 \Sigma K_1' J_0 \right]  \ \ .
\ee
If we make use of the identities
\be
J_0' &=& -J_1 \ , \nonumber \\
K_1'(z) &=& -K_0(z) -\frac 1z K_1(z) \ ,
\ee
we discover that the terms proportional to $K_1 J_0$ cancel.  This 
leaves us with
\be
G(z) = i 2 n z \Sigma^2 (1-t^2)(K_1J_1 +iK_0J_0) \ .
\ee
Finally, we can explicitly separate the resolvent into its real and imaginary 
parts by using two more elementary identities:
\be
K_1(-iz) &=& -\frac{\pi}{2} [J_1(z) + iN_1(z)] \ , \nonumber \\
K_0(-iz) &=& \frac{\pi}{2} i[J_0(z) + iN_0(z)] \ .
\ee
The final result for the microscopic spectral density is thus 
\be
\rho(\lambda) =  2 n^2 \lambda 
\Sigma^2 (1-t^2)(J_1^2(2n\lambda \Sigma \sqrt{1-t^2})
+J_0^2(2n\lambda \Sigma \sqrt{1-t^2})).
\label{specden}
\ee

As noted above, the microscopic limit of this model has previously been 
considered for the special case $t=0$ \cite{SVZr,V2,MIT,Zee}.  Our result is in 
agreement with this earlier work.   Now, however, we can also consider 
the microscopic limit for general $t \ne 0$.  At finite temperature, the 
temperature enters {\em only\/} through the temperature-dependent 
modification of the chiral condensate which was obtained in \cite{JV}.
As defined in (\ref{microscopic}) with $\rho(0)$ given by (\ref{rho(0)}), 
the microscopic limit is strictly independent of the temperature: 
\be
\rho_S (u) =  \frac u2 [J_0^2(u) + J_1^2(u)] \ .
\ee

\vskip 1.5cm
\section{Conclusions}
\vskip 0.5cm

In this paper, we have studied a random matrix model with the 
chiral structure of the QCD Dirac operator and a temperature dependence 
characteristic of the lowest Matsubara frequency.  This model possesses
the global color and flavor symmetries of QCD.  It undergoes a chiral phase
transition with critical exponents given by mean field theory. 

Using the supersymmetric method for random matrix theories, we have found
an exact, analytic expression for the average spectral density of this model.
The result has the form of a two-dimensional integral which is valid for 
matrices of any dimension.  In the large-$n$ limit, these integrals can be 
performed using a saddle-point approximation.  The spectral density then 
follows from the solution of an elementary cubic equation and nicely confirms 
our earlier numerical work \cite{JV}.

Our primary result is that the spectral density in the microscopic limit 
is strictly independent of the temperature below the critical temperature 
of this model.  This result supports the recent work of Br{\'e}zin, Hikami 
and Zee, who investigated several families of random matrix models and found 
the same microscopic limit of the spectral density in all cases.  
As noted in the introduction, our model differs from the models considered by
these authors in an essential way.  In each of their models, the spectrum of 
each element in the ensemble is strictly invariant under 
the unitary symmetry of the probability distribution. In the present model
this symmetry is violated for $T\ne 0$.
Thus, agreement between the microscopic limit of the spectral
density for our model and the models of Br{\'e}zin, Hikami 
and Zee increases our confidence in the universality of this quantity.
  
In lattice QCD simulations 
the microscopic limit of the spectral density enters in the valence quark mass
dependence of the chiral condensate. This quantity has been 
calculated for a variety of temperatures \cite{Christ}, and it has been 
shown that the results below the critical temperature and not too large
valence quark masses fall on a universal curve that
can be obtained from the microscopic limit
of the spectral density \cite{Vcr}. The present work provides 
a proper theoretical foundation of this analysis.

In our derivations, the symmetries of the partition function played a 
crucial role.  The universal behavior was closely related to the existence of
an invariant saddle-point manifold generated by these symmetries.
This suggests that the `miraculous' cancellation of the temperature dependence
of the microscopic spectral density found here is not a coincidence.  It 
would be very interesting to obtain this result using more general 
arguments.  Recent work by Guhr \cite{Guhr2} on the superposition of two 
matrix ensembles appears to offer a promising method towards this 
goal.  Work in this direction is in progress.

\vglue 0.6cm
{\bf \noindent  Acknowledgements \hfil}
\vglue 0.4cm
 The reported work was partially supported by the US DOE grant
DE-FG-88ER40388. 
\vskip 1.5cm
\noindent{\large\bf Appendix A: Notations and conventions}

In this appendix we summarize our notations and conventions.  For a more 
detailed discussion regarding the motivation for these conventions, we 
refer to \cite{Efetov,SVZ}.

The integration measure for complex Gaussian integrals is defined
such that
\be
\int \frac {d\phi^* \ d\phi}{2 \pi} \exp+i\phi^* \phi = 1 \ .
\ee
For Grassmann integrals, the measure is defined so that 
\be
2\pi  \int d\chi^* \ d\chi \exp+i\chi^* \chi = 1 \ .
\ee

A graded vector or supervector is defined by
\be
\Phi = \left ( \begin{array}{c} \phi \\ \chi \end{array}\right ) \ ,
\ee
with $\phi$ a commuting vector of length $n$ and $\chi$ an anti-commuting
vector of dimension $m$.  The corresponding supermatrix which acts on 
this vector has the structure
\be
A = \left ( \begin{array}{cc} a & \sigma \\ \rho & b \end{array} \right ) \ ,
\ee
where $a$ and $b$ are complex matrices of dimension $n \times n$ and $m \times 
m$, respectively. The entries in the $n\times m$ dimensional matrix
$\sigma$ and the $m \times n$ dimensional matrix $\rho$ are Grassmann 
variables.  The graded trace of the matrix $A$ is defined as 
\be
{\rm Trg} A = {\rm Tr} a - {\rm Tr } b \ .
\ee
The Hermitean conjugate of $A$ is defined as 
\be
A^\dagger = \left ( \begin{array}{cc} a^\dagger & \rho^\dagger 
\\ -\sigma^\dagger & b^\dagger \end{array} \right ) \ ,
\ee
where the $\dagger$ denotes transposition and complex conjugation.
A graded matrix is called Hermitean if $A^\dagger = A$.  We use 
complex conjugation of the second kind for Grassmann variables, {\em i.e.}, 
$\chi^{**} = -\chi$.  The graded determinant is defined as 
\be
{\rm detg} \ A = \exp({\rm Trg} \log A) \ .
\ee
With this definition, we obtain the following natural result for 
a Hermitean, graded matrix:  
\be
\int \prod_{i=1}^n d[{\phi}_i^*] \ d[{\phi}_i] \ 
d[{\chi}_i^*] \ d[{\chi}_i] \ \exp+i\Phi^* A \Phi 
= \frac{1}{{\rm detg} \, A}.
\label{supergauss}
\ee

\vskip 1.5cm
\noindent{\large\bf Appendix B: 
Perturbative evaluation of the average spectral density}
\vskip 0.5cm

In this appendix, we derive the large-$n$ limit of the resolvent without 
employing the supersymmetric method.  Because the operator (\ref{matrix}) 
has only a finite support, it is possible to expand the resolvent in a 
geometric series in $1/(z-K)$ for $z$ sufficiently large.  Here, $K$ is 
the matrix
\be
K = \left ( \begin{array}{cc} 0 & \pi T \\ \pi T & 0  \end{array} \right ) \ .
\ee
One finds by inspection that ${G(z)}$ satisfies
\be
G(z) = {\rm Tr}\frac 1{z-K} + {\rm Tr}\frac 1{z-K}
\overline{
\left ( \begin{array}{cc} 0 & W \\W^\dagger & 0 \end{array} \right )
{\cal G}
\left ( \begin{array}{cc} 0 & W \\W^\dagger & 0 \end{array} \right )}
{\cal G}
\ee
where ${\cal G}$ is the matrix
\be
{\cal G} = \overline{\frac 1{z-H}} \ ,
\ee
and the bar denotes averaging over the probability distribution 
(\ref{probability}).  It should be clear that ${\cal G}$ is block 
diagonal with the block structure
\newcommand{\id}{\bf 1}
\be
{\cal G} = \left ( 
\begin{array}{cc} g{\id}_n & h{\id}_n \\ h{\id}_n & g{\id}_n 
\end{array} \right ), 
\ee
where $\id_n$ is the $n\times n$ identity matrix. 
Therefore, we find that $G(z) = g$.  The average over $W$ can be carried out
immediately to give
\be
\overline{
\left ( \begin{array}{cc} 0 & W \\W^\dagger & 0 \end{array} \right )
{\cal G}
\left ( \begin{array}{cc} 0 & W \\W^\dagger & 0 \end{array} \right )}
= \frac 1{n\Sigma^2}\left (
\begin{array}{cc} g{\id}_n & 0 \\0& g{\id}_n \end{array} \right ) \ .
\ee
This yields the following matrix equation for $g$ and $h$:
\be
\left ( \begin{array}{cc} z & - \pi T \\ - \pi T & z \end{array} \right )
\left ( \begin{array}{cc} g & h\\h & g
\end{array} \right ) = {\bf 1} +\frac 1{\Sigma^2}
\left ( \begin{array}{cc} g & 0 \\0 & g
\end{array} \right )
\left ( \begin{array}{cc} g& h\\h& g
\end{array} \right ) \ , 
\ee
which leads to the two independent equations 
\be
z g - \pi T h & = & 1 + \frac 1{\Sigma^2} g^2 \ , \nonumber\\ 
z h - \pi T g & = & \frac 1{\Sigma^2} g h \ .
\ee
Elimination of $h$ yields the equation
\be
z g - \frac{\pi^2 T^2 g}{z-g/\Sigma^2} = 
1 + \frac 1{\Sigma^2} g^2 \ ,
\ee
which agrees with (\ref{spe}).  Evidently, it can be rewritten as a 
cubic equation for $g$.

\vfill
\eject
\newpage
\setlength{\baselineskip}{14pt}

\end{document}